\documentclass[english,aps,prd,superscriptaddress,nofootinbib,eqsecnum,showpacs,10pt,letterpaper]{revtex4-1}
\usepackage[T1]{fontenc}
\usepackage{amsmath}
\usepackage{graphicx}
\usepackage{amssymb}

\usepackage[dvips]{color}

\makeatletter

\def\be{\begin{equation}}\def\ee{\end{equation}}\def\te{\end{equation}}\def\bea{\begin{eqnarray}}\def\ba{\begin{eqnarray}}\def\ta{\end{eqnarray}}\def\tea{\end{eqnarray}}

\def\ben{\begin{enumerate}}\def\een{\end{enumerate}}

\def\e{\epsilon}

\makeatother

\makeatother

\usepackage{babel}

\begin{document}

\title{Autangle: A case of Quantum Narcissism?}

\author{Rong Zhou}
\email{zhour@umd.edu}
\affiliation{Joint Quantum Institute and Maryland Center for Fundamental
Physics,
University of Maryland, College Park, Maryland 20742, USA}

\author{Ryan O. Behunin}
\email{rbehunin@lanl.gov}
\affiliation{Center for Nonlinear Studies and Los Alamos National Laboratory,
Theoretical Division, Los Alamos, New Mexico 87545, USA}

\author{Shih-Yuin Lin}
\email{sylin@cc.ncue.edu.tw}
\affiliation{Department of Physics, National Changhua University of Education,
Changhua 50007, Taiwan}

\author{Bei-lok Hu}
\email{blhu@umd.edu}
\affiliation{Joint Quantum Institute and Maryland Center for Fundamental
Physics,
University of Maryland, College Park, Maryland 20742, USA}

\date{\today}
\begin{abstract}In this paper we ask a common psychological question and provide a
physics answer: {}``Looking into a mirror can one get entangled with
one's image?\textquotedbl{} This is not a frivolous question; rather, it bears on the effect of
boundaries on the behavior of quantum entanglement  between a harmonic oscillator and a quantum field,  a basic problem of interest in proposed mirror-field superposition and related experiments in macroscopic quantum phenomena, as well as atomic fluctuation forces near a conducting surface. The object's internal degree of freedom is modeled by a harmonic oscillator and the presence of a perfectly reflecting mirror enforces the Dirichlet boundary conditions on the quantum field, restricting the latter to a half space.
By assuming a bilinear oscillator-field interaction, we derive a coupled set of equations for the oscillator's and the field's Heisenberg operators. 
The former can be cast in the form of a quantum Langevin equation, where the dissipation and noise kernels respectively correspond to the retarded and Hadamard functions of the free quantum field in half space.
 We use the linear entropy as measures of entanglement between the oscillator and the quantum field under mirror reflection, then solve the early-time oscillator-field entanglement dynamics and compare it with that between two inertial oscillators in free space. 
At late times when the combined system is in a stationary state, we obtain exact expressions for the oscillator's covariance matrix and show that the oscillator-field entanglement decreases as the oscillator moves closer to the mirror. We explain this behavior qualitatively with the help of a mirror image and provide an answer to the question raised above. 
 We also compare this situation with the case of two real oscillators and explain the differences.
\end{abstract}

\maketitle

\section{Introduction}

In this note we ask a common psychological question and provide a
physics answer: {}``Looking into a mirror can one get entangled with
one's image\textquotedbl{}? According to storybooks the answer for
the evil queen Q seems to be yes, a case of typical royal narcissism.
We want to find out the answer from physical considerations: here
entanglement refers to quantum entanglement, and \textit{autangle}
means self-entanglement 
\footnote{We coin this term from two sources of inspiration: 
new words like 3-tangle and old ones like autism. A more learned word `ipso-tangle' was suggested by Prof. Brill.}
,
 here referring to the entanglement of a real
physical object Q with its mirror image. We hasten to add that since
the image is not a physical object one cannot define, let alone calculate,
the entanglement between a physical object and an unphysical construct.
However, there exists entanglement between an oscillator and a quantum field,
and we can ask how this is altered if a mirror is present. A perfectly
reflecting mirror imposes Dirichlet boundary conditions on the field
along the mirror surface, restricting its existence to a half space.
These are well-defined problems, belonging to the broader inquiry
into the effects of boundaries and topology on the quantum field and
on quantum entanglement of  objects, oscillators and mirrors, coexisting
with and/or mediated by the quantum field. In fact, they bear on issues
of importance to quantum information and macroscopic quantum phenomena.
Amongst research problems of current interest, we mention: 1) Entanglement
between two two-level (2LA) atoms via a common quantum field already shows diverse quantum entanglement
dynamics behavior \cite{SCH1}, ranging from sudden death, touch of death, revival
to staying always alive, and features such as dynamical generation,
protection, and transfer of entanglement between subsystems. Model
studies of oscillator-field entanglement with experimental ventures have
been carried out \cite{Sabrina} as well as oscillator-field entanglement
in conjunction with measurements in LIGO detectors \cite{Miao}, the
latter also serving as preparatory studies for mirror-field superposition
\cite{Marshall} in macroscopic quantum phenomena. 2) Entanglement
not only changes in time but it also depends on spatial separation
as shown in model studies for 2LAs \cite{ASH,FCAH_2atom,SCH2} and harmonic
oscillators \cite{LH_2IHO} interacting with a common quantum field.
These results are of both theoretical interest in understanding what
quantum nonlocality means, and practical value, such as for the design
of quantum gates and applications to quantum teleportation. Extending
these findings to two and three dimensional systems would enable one
to define quantum entanglement domains \cite{ZFHed,Edomain} and geometric effects of entanglement
\cite{ZFHep,Epattern}. The former refers to the effective domain
on a surface or body where entanglement is induced by the presence
of an oscillator, not unlike the induced surface charge density on a conductor
or dielectric plate. The latter refers to specific geometric patterns
in the arrangements of oscillators where their entanglement strength can
be maximized.

Now, returning to the problem under study, one may wonder, is it really
necessary to carry out such a serious calculation in order to answer
a simple query posed at the beginning: Is there entanglement between
the queen and her image? The answer is yes, both to the query and
to the necessity of a bona fide calculation. As we cautioned {\it ab initio}, an image is not a physical
object. Thus we try to avoid invoking an image in this calculation
so that no unphysical assumptions or intuition are brought in unnecessarily
or unknowingly in the derivations. However, at the end, by inspecting
the results from the oscillator-field (under mirror reflection) entanglement
we see that one can use the notion of an image to describe the process,
in fact, come up with a simple qualitative explanation of how it depends
on distance.

The paper is organized as follows: In Section. II we set up the
problem and derive a formal set of equations for the Heisenberg operators
of the oscillator's internal degrees of freedom and the quantum field. We
impose the boundary conditions introduced by the presence of the mirror
and explain how the commutation relations are altered. Next we calculate
the retarded and Hadamard Green functions of the altered field configurations
and derive a quantum Langevin equation for $Q(t)$ including the back-reaction
of the altered field. We seek solutions to this equation at late times
and calculate the covariance matrix of the oscillator's canonical variables
at late times when the combined system is in a stationary state. Details
are contained in Appendix \ref{LateCV}.
In Sec. III we introduce the linear entropy and the von-Neumann entropy as measures of quantum entanglement
and calculate the early-time dynamics of entanglement between the oscillator and
the field in the oscillator-mirror setup described above.
We obtain plots of how the entanglement between the oscillator and field in
the half space with the mirror reflection evolves with time, depending on the distance
between the oscillator and the mirror.
Finally in the discussions section
we explain in what sense can one describe this situation in terms
of entanglement of the oscillator with its mirror image and why it decreases
as the oscillator moves closer to the mirror, a somewhat counterintuitive
finding. We also compare this situation with the case of two inertial
oscillators in free space (calculations placed in Appendix \ref{w2IHO}) and explain
their physical differences.

\section{Oscillator Interacting with Field under Mirror Boundary
Condition}

\subsection{Dynamics of Oscillator-Field System with Mirror}

As described in the Introduction, the gist of the matter for this
problem is to quantify the change in some entanglement measure between
the oscillator and the field with and without the presence of a mirror.
Technically our calculation involves two parts: 1) find the back-reaction
of the field configuration altered by the mirror on the oscillator. 2) find
the correlation functions of the resultant oscillator's canonical variables including the field's influences.
Thus our target is the covariance matrix for the oscillator which incorporates
all back reactions from the quantum field. We work with the Heisenberg
equations of motion which efficiently provides the oscillator's field-influenced
dynamics. We can then find the oscillator's late time behavior and quantify
it's entanglement with the modified field.

Consider an oscillator located at position ${\bf x}_Q^{}$ at a vertical distance of $L/2$ from the mirror plane at $x_3=0$. The total action of the system describing this oscillator with internal degree of freedom $Q$ interacting linearly with a massless scalar field confined to the half space defined by $x_{3}>0$ is given by:
\begin{equation}
S[Q,\dot{Q};\Phi,\partial_{\mu}\Phi]=\frac{1}{2}M_{Q}\int dt\ (\dot{Q}^{2}-\Omega^{2}Q^{2})+\frac{1}{2}\int dt\int_{x_{3}>0}d^{3}x\ \partial_{\mu}\Phi\partial^{\mu}\Phi+\lambda_{Q}\int dt\ Q(t)\Phi({\bf x}_Q^{},t).\label{eq:Action_Half}\end{equation}
 Due to the linearity of our system the equations of motion for Heisenberg
operators (carrying hats) have the same form as the equations for
the corresponding classical variables:
\begin{align}
M_{Q}\ddot{\hat{Q}}(t)+M_{Q}\Omega^{2}{\hat{Q}}(t)= & \lambda_{Q}\int_{x_{3}>0}d^{3}x\ {\hat{\Phi}}({\bf x}_Q^{},t),
\label{AtomEqn}\\
\Box{\hat{\Phi}}({\bf x},t)= & \lambda_{Q}\delta^{3}({\bf x}-{\bf x}_Q^{}){\hat{Q}}(t),
\quad{\rm and}\quad{\hat{\Phi}}({\bf x}_{\|},x_{3}=0,t)=0.\label{FieldEqn}\end{align}
Equation (\ref{FieldEqn}) specifies the Dirichlet boundary (DBC)
conditions imposed on the field where the mirror surface is located,
namely, in the $x_{3}=0$ plane and ${\bf x}_{\|}=(x_1,x_{2})$. To
obtain the back-reaction of the field on the oscillator we first solve (\ref{FieldEqn})
and then plug the solution into (\ref{AtomEqn}). The solution for
${\hat{\Phi}}({\bf x},t)$ is given by
\begin{equation}
{\hat{\Phi}}({\bf x},t)={\hat{\Phi}}_{0}({\bf x},t)+\lambda_{Q}\int_{t_{i}}^{t}dt'G_{ret}^{\Phi}(t,{\bf x};t',{\bf x}_Q^{}){\hat{Q}}(t')\label{FieldSoln}\end{equation}
 where ${\hat{\Phi}}_{0}({\bf x},t)$ is the homogeneous solution
to equation (\ref{FieldEqn}) which describes the dynamics of the
source-free field (without $Q$) in the presence of the mirror and
$G_{ret}^{\Phi}({\bf x},t;{\bf y},t^{\prime})$ is the retarded Green's
function for the field. Plugging the solution for the field operator
into the equation of motion for the oscillator we obtain the following
equation governing the dynamics of the oscillator with the effects of the
field already incorporated. This is what we mean by the `field-influenced'
dynamics of the oscillator
\begin{align}
M_{Q}\ddot{\hat{Q}}(t)+M_{Q}\Omega^{2}{\hat{Q}}(t)-\lambda_{Q}^{2}\int_{t_{i}}^{t}dt'G_{ret}^{\Phi}(t,{\bf x}_Q^{};t',{\bf x}_Q^{}){\hat{Q}}(t')=\lambda_{Q}{\hat{\Phi}}_{0}({\bf x}_Q^{},t).\label{FI-AtomEqn}\end{align}

As will be explained in further detail later the term containing the
retarded Green's function describes how the oscillator transfers energy
to the field and the right hand side acts similarly to a Langevin
forcing term describing how quantum field fluctuations drive the oscillator.

\subsection{Field Modified by Mirror and Propagators in Half-space}

In order to solve the equation of motion for the oscillator (\ref{FI-AtomEqn})
we need to know the field operators defined in the half space. Following
this procedure we first write down the free field operator $\hat{\Phi}_{0}$
which satisfies the Klein-Gordon equation and vanishes in the $x_{3}=0$
plane,
\begin{equation}
{\hat{\Phi}}_{0}({\bf x},t)=\int_{k_{3}>0}d^{3}k\sqrt{\frac{1}{4\pi^{3}\omega}}
e^{-i\omega t+i{\bf k}_{\|}\cdot {\bf x}_{\|}}\sin k_{3}x_{3}{\hat{b}}_{{\bf k}}+H.c.\label{ }
\end{equation}
where $\omega=|{\bf k}|$.

Notice that the normalization factor is different for field in half space.

Writing ($\hat{\Pi}_{0}({\bf x}',t)=\dot{\hat{\Phi}}_{0}({\bf x}',t)$)
and demanding that the equal-time commutation relations are satisfied,
that is,
\begin{equation}
[\hat{\Phi}_{0}({\bf x},t),\hat{\Pi}_{0}({\bf x}',t)]=i\delta^{(3)}({\bf x}-{\bf x}')-i\delta^{(2)}({\bf x}_{\|}-{\bf x}_{\|}')
\delta(x_{3}+x_{3}'),\label{CommRel}
\end{equation}
we can infer that the mode expansion coefficients can be identified
with the creation and annihilation operators
\begin{equation}
[{\hat{b}}_{{\bf k}},{\hat{b}}_{{\bf k}}^{\dag}]=\delta_{{\bf k},{\bf k}'},
\quad[{\hat{b}}_{{\bf k}},{\hat{b}}_{{\bf k}}]=0,
\quad[{\hat{b}}_{{\bf k}}^{\dag},{\hat{b}}_{{\bf k}}^{\dag}]=0.
\end{equation}
 Note that the commutation relation above (\ref{CommRel}) is not
in the form normally encountered. For the case of a field satisfying
the Dirichlet boundary conditions on the $x_{3}=0$ plane it is necessary
to add the \textquotedbl{}image term\textquotedbl{} to the right hand
side 
such that for points restricted to the $x_3<0$ half space where the
image term has no support the relation (\ref{CommRel}) reduces to
the standard form.

With the expression for the free field operator in hand we now calculate
the retarded and Hadamard propagators. The retarded propagator quantifies
the field sourced by the oscillator and describes how energy is dissipated
from the oscillator into the field
\begin{align}
G_{ret}^{\Phi}({\bf x},t;{\bf y},t^{\prime}) & \equiv i\theta(t-t^{\prime})[\hat{\Phi}_{0}({\bf x},t),\hat{\Phi}_{0}({\bf y},t')]\nonumber \\
 & =\theta(t-t^{\prime})\frac{1}{2\pi^{3}}\int_{k_{3}>0}d^{3}k\frac{1}{\omega}\sin(\omega(t-t^{\prime}))e^{i{\bf k}_{\parallel}\cdot({\bf x}_{\parallel}-{\bf y}_{\parallel})}\sin(k_{3}x_{3})\sin(k_{3}y_{3})\\
 & =\theta(t-t^{\prime})\int_{0}^{\infty}d\omega\sin(\omega(t-t^{\prime}))\cdot I(\omega;{\bf x},{\bf y}).\nonumber \end{align}
 The Hadamard function, which will be encountered in the next section,
describes the quantum fluctuations of the field and is given formally by the anti-commutator of the field operator:
\begin{align}
G_{H}^{\Phi}({\bf x},t;{\bf y},t^{\prime}) & \equiv\langle\{\hat{\Phi}_{0}({\bf x},t),\hat{\Phi}_{0}({\bf y},t')\}\rangle\nonumber \\
 & =\frac{1}{2\pi^{3}}\int_{k_{3}>0}d^{3}k\frac{1}{\omega}\cos(\omega(t-t^{\prime}))e^{i{\bf k}_{\parallel}\cdot({\bf x}_{\parallel}-{\bf y}_{\parallel})}\sin(k_{3}x_{3})\sin(k_{3}y_{3})\\
 & =\int_{0}^{\infty}d\omega\cos(\omega(t-t^{\prime}))\cdot I(\omega;{\bf x},{\bf y}),\nonumber \end{align}
 where \begin{equation}
I(\omega;{\bf x},{\bf y})=\frac{\omega}{2\pi^{3}}\int_{0}^{\pi/2}d\theta\int_{0}^{2\pi}d\phi\ e^{i{\bf k}_{\|}\cdot({\bf x}_{\parallel}-{\bf y}_{\parallel})}\sin(\omega\cos\theta x_{3})\sin(\omega\cos\theta y_{3}).\end{equation}

\section{Early Time Dynamics of Oscillator-Field Entanglement}

In \cite{LH_2IHO} the dynamics of the entanglement between two inertial oscillators
with identical couplings to a common massless scalar field was investigated
and it exhibited several interesting features.
There are roughly three temporal regimes if the two oscillators are properly separated
and weakly coupled to the field. The first regime is at early time, up to $t\approx O(1/\lambda_Q^2)$
 when accumulated effects of field-mediated mutual
influences, which manifest themselves in the field's retarded propogators, between the oscillators
are still weak and the reduced dynamics of the oscillators are dominated by influences of vacuum fluctuations of the field.
As the effects of field-mediated mutual influences between the oscillators gradually gain strength, the system will enter an intermediate
regime in which the oscillators' reduced dynamics become quite complicated.
In the late-time limit, the oscillator approaches 
a time-stationary state.

In particular, in \cite{LH_2IHO} it was shown that at early time, after causal contact has been established
between two detectors,
an oscillatory pattern of entanglement emerges which varies with the distance between the oscillators.
This is not due to mutual influences since in the weak coupling limit the mutual influence effect 
is always weak compared to effect of vacuum fluctuation.

For our model, we expect a similar division into three temporal regimes.
Here in this section we study the early time behavior of our model, assuming
that initially both the oscillator and the field are in their ground states
and are uncorrelated. We show that the oscillator-field entanglement also
develops an oscillatory pattern which varies with the distance
between the oscillator and the mirror. At a given instant of time, the oscillator-field
entanglement exhibits spatial oscillations characterized by frequency
$\Omega$ (we adopt the convention $c=1$).
Additionally, we find that the growth rate at which the oscillator-field become entangled
oscillates as a function of the oscillator-mirror spacing.

\subsection{Measure of Oscillator-Field Entanglement }

In our setup, the oscillator and the field together form a closed quantum
system which undergoes unitary evolution. Since the initial state
of the total system is Gaussian and the action is quadratic, the reduced
quantum state of the oscillator at any time will remain Gaussian. For such systems the amount
of bipartite entanglement between the oscillator and the field can be quantified
by the linear entropy of the oscillator's reduced density matrix, which takes
on values different from $0$ when the oscillator and field are entangled.
Here the linear entropy is defined as
\begin{equation}
S_{L}=1-\mathcal{P},\end{equation}
where $\mathcal{P}\triangleq Tr\hat{\rho}_{\rm a}^{2}$ is the purity of
the oscillator's reduced quantum state. Here for our general mixed Gaussian
state, the purity is given by
\begin{equation}
\mathcal{P}=\frac{1}{2\sqrt{\det V}}=\frac{1}{2\sqrt{<\hat{Q},\hat{Q}><\hat{P},\hat{P}>-<\hat{Q},\hat{P}>^{2}}}\end{equation}
where $<\hat{O}_{i},\hat{O}_{j}>\triangleq Tr(\hat{\rho}_{\rm a}\cdot\{\hat{O}_{i},\hat{O}_{j}\})/2$,
 $\boldsymbol{\hat{O}}=(\hat{Q},\hat{P})$.

Mathematically we know that $\mathcal{P}$ is always less than or equal to 1. Smaller purity means the reduced density matrix of the oscillator
is less pure, and correspondingly the oscillator is more entangled with the field.

Another measure of bipartite entanglement between the oscillator and the
field is von Neumann entropy, which for a single-mode mixed Gaussian
state is
\begin{equation}
S_{\nu}=\frac{1-\mathcal{P}}{2\mathcal{P}}\ln(\frac{1+\mathcal{P}}{1-\mathcal{P}})-\ln(\frac{2\mathcal{P}}{1+\mathcal{P}})\end{equation}
and is a monotonically
increasing function of the linear entropy. Both $S_{L}$ and $S_{\nu}$ yield
the same characterization of mixedness and are equivalent as entanglement
measures in the case of a single-mode mixed Gaussian state.

\subsection{Mode Decomposition and EOM for Mode Functions}

We perform the following mode decompositions for Heisenberg operators
of the oscillator $\hat{Q}(t)$ and the field $\hat{\Phi}(x)$ with
respect to the creation and annihilation operators of the oscillator
$\hat{a},\hat{a}^{\dagger}$ and those of the field $\hat{b},\hat{b}^{\dagger}$
:

\begin{eqnarray}
\hat{Q}(t) & = & \sqrt{\frac{\hbar}{2\Omega_{r}}}\left[q_{\rm a}(t)\hat{a}+q_{\rm a}^{*}(t)\hat{a}^{\dagger}\right]+\int_{k_{3}>0}\frac{d^{3}k}{\sqrt{2\pi^{3}}}\sqrt{\frac{\hbar}{2\omega}}\left[q_{+}(t,{\bf k})\hat{b}_{{\bf k}}+q_{-}(t,{\bf k})\hat{b}_{{\bf k}}^{\dagger}\right],\\
\hat{\Phi}(x) & = & \sqrt{\frac{\hbar}{2\Omega_{r}}}\left[f_{\rm a}(x)\hat{a}+f_{\rm a}^{*}(x)\hat{a}^{\dagger}\right]+\int_{k_{3}>0}\frac{d^{3}k}{\sqrt{2\pi^{3}}}\sqrt{\frac{\hbar}{2\omega}}\left[f_{+}(x;{\bf {\bf k}})\hat{b}_{{\bf k}}+f_{-}(x;{\bf {\bf k}})\hat{b}_{{\bf k}}^{\dagger}\right].\end{eqnarray}

By plugging (3.4) and (3.5) into (2.2) and (2.3), respectively, we find the following equations of motion for the mode functions

\begin{eqnarray}
\left(\partial_{t}^{2}+2\gamma_Q^{}\partial_{t}+\Omega_{r}^{2}\right)q_{\rm a}(t) & =- & \frac{2\gamma_Q^{}}{L}\theta(t-L)q_{\rm a}(t-L),\label{EOMqq}\\
\left(\partial_{t}^{2}+2\gamma_Q^{}\partial_{t}+\Omega_{r}^{2}\right)q_{+}(t,{\bf k}) & = & -\frac{2\gamma_Q^{}}{L}\theta(t-L)q_{+}(t-L,{\bf k})+{\lambda_{Q}\over M_Q}f_{0+}(t,{\bf x}_Q^{};{\bf k}),
\label{EOMq+}\end{eqnarray}
in which $f_{0+}(t,{\bf x}_Q^{};{\bf k})=e^{-i\omega t+i{\bf k}_{\parallel}\cdot{\bf x}_{\parallel}}\sin k_{3}x_{3}$
and $\gamma_Q^{}\equiv \lambda_Q^2/(8\pi M_Q)$.

Since we have assumed the oscillator's position to be 
${\bf x}_Q^{}(t)=\bar{\bf x}_Q^{} \equiv (0,0,L/2)$, we have $f_{0+}(\bar{\bf x}_Q^{},t;{\bf k})=e^{-i\omega t}\sin\frac{k_{3}L}{2}$.

According to our initial condition, the solutions have to satisfy
initial conditions that
$f_{+}(0,{\bf x};{\bf k})= e^{i {\bf k}_{\parallel}\cdot{\bf x}_{\parallel}}\sin k_{3}x_{3}$, 
$\partial_{t}f_{+}(0, {\bf x};{\bf k})=-i\omega e^{i{\bf k}_{\parallel}\cdot{\bf x}_{\parallel}}\sin k_{3}x_{3}$, 
$q_{\rm a}(0)=1$, $\partial_{t}q_{\rm a}(0)=-i\Omega_{r}$,
and $f_{\rm a}(0,{\bf x})=\partial_{t}f_{\rm a}(0,{\bf x})=q_{+}(0;{\bf k})=\partial_{t}q_{+}(0;{\bf k})=0$.

The solution to (\ref{EOMqq}) can be written as the following expansion
which is truncated at $nL>t$,

\begin{equation}
q_{q}(t)=\sum_{n=0}q_{q}^{(n)}(t),\label{qexpan}\end{equation}
 where\begin{align}
q_{\rm a}^{(0)}(t) & =q_{\rm a}^{(h)}(t),\\
q_{\rm a}^{(n)}(t) & =\int d\tau_{1}G_{r}(t,\tau_{1})(-\frac{2\gamma_Q^{}}{L})\theta(t-L)\times\\
 & ...\int d\tau_{n}G_{r}(\tau_{n-1}-\tau_{n})(-\frac{2\gamma_Q^{}}{L})\theta(\tau_{n}-nL)q_{\rm a}^{(h)}(\tau_{n}-nL).\nonumber \end{align}
Here $q_{\rm a}^{(h)}(t)$ is the homogeneous solution satisfying the initial
conditions

\begin{equation}
q_{\rm a}^{(h)}(t)=\frac{1}{2}(1+\frac{\Omega_{r}+i\gamma_Q^{}}{\Omega})e^{-\gamma_Q^{} t-i\Omega t}+\frac{1}{2}(1-\frac{\Omega_{r}+i\gamma_Q^{}}{\Omega})e^{-\gamma_Q^{} t+i\Omega t}.\end{equation}

The solution to (\ref{EOMq+}) can be written in a similar fashion as above, also truncated at $nL>t$,

\begin{equation}
q_{+}(t,{\bf k})=\sum_{n=0}q_{+}^{(n)}(t,{\bf k}),\label{pexpan}\end{equation}
 where

\begin{align}
q_{+}^{(0)}(t,{\bf k}) & =\int d\tau_{1}G_{r}(t,\tau_{1})\lambda_{Q}f_{0+}(\tau_{1},\bar{\bf x}_Q^{};{\bf k})\\
 & =\frac{\lambda_{Q}}{M_Q\Omega}
 \left(\sin\frac{k_{3}L}{2}\right)[(M_{1}-M_{2})e^{-i\omega t}+(M_{2}e^{i\Omega t}-M_{1}e^{-i\Omega t})e^{-\gamma_Q^{} t}],\nonumber \\
q_{+}^{(n-1)}(t,{\bf k}) & =\int d\tau_{1}G_{r}(t,\tau_{1})(-\frac{2\gamma_Q^{}}{L})\theta(\tau_{1}-L)\times\\
 & \int d\tau_{2}G_{r}(\tau_{1},\tau_{2})(-\frac{2\gamma_Q^{}}{L})\theta(\tau_{2}-2L)\times\nonumber \\
 & ...\int d\tau_{n}G_{r}(\tau_{n-1}-\tau_{n})\theta(\tau_{n}-(n-1)L)
 \lambda_{Q}\over M_Q
 f_{0+}(\tau_{n}-(n-1)L,\bar{\bf x}_Q^{};{\bf k}),\nonumber \end{align}

where $M_{1}=1/[2(-\omega-i\gamma_Q^{}+\Omega)]$, $M_{2}=1/[2(-\omega-i\gamma_Q^{}-\Omega)]$,
$G_{r}(t,\tau)$ is the retarded Green's function which satisfies
$(\partial_{t}^{2}+2\gamma_Q^{}\partial_{t}+\Omega_{r}^{2})G_{r}(t,\tau)=\delta(t-\tau)$.

\subsection{Zeroth-Order Correlation Functions}

As explained before, in order to understand the dynamics of entanglement,
it is sufficient to compute the covariance matrix of the oscillator's reduced
quantum state. For a separable initial state, the covariance
matrix can be decomposed into two parts corresponding to the two sets
of operators in the mode decomposition:

\begin{eqnarray}
<\hat{Q}(t),\hat{Q}(t)> & = & <\hat{Q}(t),\hat{Q}(t)>_{\rm a}+<\hat{Q}(t),\hat{Q}(t)>_{\rm v}\\
 & = & \frac{1}{2\Omega_{r}}|q_{\rm a}(t)|^{2}+\int_{k_{3}>0}\frac{d^{3}k}{2\pi{}^{3}}\frac{1}{2\omega}|q_{+}(t,{\bf k})|^{2},\nonumber \end{eqnarray}

\begin{eqnarray}
 & < & \hat{P}(t),\hat{P}(t)>=<\hat{P}(t),\hat{P}(t)>_{\rm a}+<\hat{P}(t),\hat{P}(t)>_{\rm v}\\
 & = & M_Q^2 \left[
 \frac{1}{2\Omega_{r}}|\partial_{t}q_{\rm a}(t)|^{2}+
 \int_{k_{3}>0}\frac{d^{3}k}{2\pi{}^{3}}\frac{1}{2\omega}|\partial_{t}q_{+}(t,{\bf k})|^{2}\nonumber
 \right],
\end{eqnarray}

\begin{eqnarray}
 & < & \hat{Q}(t),\hat{P}(t)>=<\hat{Q}(t),\hat{P}(t)>_{\rm a}+<\hat{Q}(t),\hat{P}(t)>_{\rm v}\\
 & = & \frac{M_Q}{2\Omega_{r}}\frac{1}{2}[q_{\rm a}^{*}(t)\partial_{t}q_{\rm a}(t)+q_{\rm a}^{*}(t)\partial_{t}q_{\rm a}(t)]+
 \int_{k_{3}>0}\frac{d^{3}k}{2\pi{}^{3}}\frac{M_Q}{2\omega}\frac{1}{2}[q_{+}^{*}(t,{\bf k})
 \partial_{t}q_{+}(t,{\bf k})+q_{+}(t,{\bf k})\partial_{t}q_{+}^{*}(t,{\bf k})].\nonumber
\end{eqnarray}
Here $<\hat{O}_i,\hat{O}_j> \equiv < (\hat{O}_i\hat{O}_j+\hat{O}_j\hat{O}_i)>/2$. 
In the weak coupling limit, the effect of reflected influences correspond to terms in \ref{qexpan} and \ref{pexpan}
 which are of higher than zeroth order, 
and therefore at early time (up to $t\approx 1/\gamma_Q $) accumulated effect of reflected influences is always small.
Thus for the purpose of studying the early time behavior of
entanglement, we can ignore the contribution of reflected influences and
restrict ourselves to lowest order correlations.

We have
\begin{eqnarray*}
 & <\hat{Q}(t),\hat{Q}(t)>_{\rm a}=\frac{1}{2\Omega_{r}}|q_{\rm a}^{(h)}(t)|^{2}\\
= & \frac{1}{2\Omega_{r}}|\frac{1}{2}(1+\frac{\Omega_{r}+i\gamma_Q^{}}{\Omega})e^{-\gamma_Q^{} t-i\Omega t}+\frac{1}{2}(1-\frac{\Omega_{r}+i\gamma_Q^{}}{\Omega})e^{-\gamma_Q^{} t+i\Omega t}|^{2},\end{eqnarray*}

\begin{eqnarray*}
 & <\hat{P}(t),\hat{P}(t)>_{\rm a}=\frac{M_Q^2}{2\Omega_{r}}|\partial_{t}q_{\rm a}^{(h)}(t)|^{2}\\
= & \frac{M_Q^2}{2\Omega_{r}}|\frac{1}{2}(1+\frac{\Omega_{r}+i\gamma_Q^{}}{\Omega})(\gamma_Q^{}+i\Omega)e^{-\gamma_Q^{} t-i\Omega t}+\frac{1}{2}(1-\frac{\Omega_{r}+i\gamma_Q^{}}{\Omega})(\gamma_Q^{}-i\Omega)e^{-\gamma_Q^{} t+i\Omega t}|^{2},\end{eqnarray*}

\begin{eqnarray*}
 & <\hat{Q}(t),\hat{P}(t)>_{\rm a}=\frac{M_Q}{2\Omega_{r}}\frac{1}{2}
 (q_{\rm a}^{(h)*}(t)\cdot\partial_{t}q_{\rm a}^{(h)}(t)+q_{\rm a}^{(h)}(t)\cdot\partial_{t}q_{\rm a}^{(h)*}(t))\\
= & \frac{M_Q}{2\Omega_{r}}Re\{[\frac{1}{2}(1+\frac{\Omega_{r}+i\gamma_Q^{}}{\Omega})
(\gamma_Q^{}+i\Omega)e^{-\gamma_Q^{} t-i\Omega t}+\frac{1}{2}(1-\frac{\Omega_{r}+i\gamma_Q^{}}{\Omega})(\gamma_Q^{}-i\Omega)e^{-\gamma_Q^{} t+i\Omega t}]\\
 & \cdot[\frac{1}{2}(1+\frac{\Omega_{r}-i\gamma_Q^{}}{\Omega})e^{-\gamma_Q^{} t+i\Omega t}+
 \frac{1}{2}(1-\frac{\Omega_{r}-i\gamma_Q^{}}{\Omega})e^{-\gamma_Q^{} t-i\Omega t}]\}.\end{eqnarray*}

Notice that the zeroth order correlators $<...>_{\rm a}^{(0)}$'s do not
depend on the distance between the oscillator and the mirror, but the  part induced by vacuum fluctuations does:

\begin{eqnarray}
 & <\hat{Q}(t),\hat{Q}(t)>_{\rm v}=\int_{k_{3}>0}\frac{d^{3}k}{2\pi{}^{3}}\frac{1}{2\omega}|q_{+}^{(0)}(t,{\bf k})|^{2} \label{QQv}\\
 &=(\frac{\lambda_{Q}}{M_Q\Omega})^{2}
  \int_{k_{3}>0}\frac{d^{3}k}{2\pi{}^{3}}\frac{1}{2\omega}(\sin\frac{k_{3}L}{2})^{2}|(M_{1}-M_{2})e^{-i\omega t}+
  (M_{2}e^{i\Omega t}-M_{1}e^{-i\Omega t})e^{-\gamma_Q^{} t}|^{2}\nonumber \\
 & =(\frac{\lambda_{Q}}{M_Q\Omega})^{2}\frac{1}{4\pi^{2}}
 \int d\omega {\omega\over 2}\left( 1-\frac{\sin\omega L}{\omega L}\right)
 |(M_{1}-M_{2})e^{-i\omega t}+(M_{2}e^{i\Omega t}-M_{1}e^{-i\Omega t})e^{-\gamma_Q^{} t}|^{2}.\nonumber \end{eqnarray}
Similarly we have

\begin{eqnarray}
 & <\hat{P}(t),\hat{P}(t)>_{\rm v}=M_Q^2
 \int_{k_{3}>0}\frac{d^{3}k}{2\pi{}^{3}}\frac{1}{2\omega}|\partial_{t}q_{+}^{(0)}(t,{\bf k})|^{2}\\
 & =(\frac{\lambda_{Q} M_Q}{\Omega})^{2}\frac{1}{4\pi^{2}}\int d\omega
 {\omega\over 2}\left( 1-\frac{\sin\omega L}{\omega L}\right)\times\nonumber \\
 & |-i\omega(M_{1}-M_{2})e^{-i\omega t}+((i\Omega-\gamma_Q^{})M_{2}e^{i\Omega t}-(-i\Omega-\gamma_Q^{})M_{1}e^{-i\Omega t})e^{-\gamma_Q^{} t}|^{2},\nonumber \end{eqnarray}

\begin{eqnarray}
 & <\hat{Q}(t),\hat{P}(t)>_{\rm v}=M_Q\int_{k_{3}>0}\frac{d^{3}k}{2\pi{}^{3}}\frac{1}{2\omega}\frac{1}{2}(q_{+}^{(0)*}(t,{\bf k})\partial_{t}q_{+}^{(0)}(t,{\bf k})+q_{+}^{(0)}(t,{\bf k})\partial_{t}q_{+}^{(0)*}(t,{\bf k}))\\
 & =  M_Q
 (\frac{\lambda_{Q}}{\Omega})^{2}\frac{1}{4\pi^{2}}
 \int d\omega {\omega\over 2}\left( 1-\frac{\sin\omega L}{\omega L}\right)\times\nonumber \\
 & Re\{[-i\omega(M_{1}-M_{2})e^{-i\omega t}+((i\Omega-\gamma_Q^{})M_{2}e^{i\Omega t}-(-i\Omega-\gamma_Q^{})M_{1}e^{-i\Omega t})e^{-\gamma_Q^{} t}]\nonumber \\
 & \cdot[(M_{1}-M_{2})e^{i\omega t}+(M_{2}e^{i\Omega t}-M_{1}e^{-i\Omega t})e^{-\gamma_Q^{} t}]\}.\nonumber \end{eqnarray}

In \cite{LH_2IHO}, for the case of two inerial oscillators distance $L$ apart,
which are located in free space with identical couplings to the field, the
zeroth order correlators due to vacuum fluctuation are given as
($M_Q \equiv 1$ and $d$ in \cite{LH_2IHO} is replaced by $L$):

\begin{eqnarray*}
 & <\hat{Q}_{A}(t),\hat{Q}_{B}(t)>_{\rm v}=(\frac{\lambda_{Q}}{\Omega})^{2}\frac{1}{4\pi^{2}}
 \int d\omega \,\omega\, \frac{\sin\omega L}{\omega L}\cdot\\
 & |(M_{1}-M_{2})e^{-i\omega t}+(M_{2}e^{i\Omega t}-M_{1}e^{-i\Omega t})e^{-\gamma_Q^{} t}|^{2},
\label{QAQBv}
\end{eqnarray*}

\begin{eqnarray*}
 & <\hat{Q}_{A}(t),\hat{P}_{B}(t)>_{\rm v}=(\frac{\lambda_{Q}}{\Omega})^{2}\frac{1}{4\pi^{2}}
 \int d\omega \,\omega\, \frac{\sin\omega L}{\omega L}\cdot\\
 & Re\{[-i\omega(M_{1}-M_{2})e^{-i\omega t}+((i\Omega-\gamma_Q^{})M_{2}e^{i\Omega t}-(-i\Omega-\gamma_Q^{})M_{1}e^{-i\Omega t})e^{-\gamma_Q^{} t}]\cdot\\
 & [(M_{1}-M_{2})e^{i\omega t}+(M_{2}e^{i\Omega t}-M_{1}e^{-i\Omega t})e^{-\gamma_Q^{} t}]\},\end{eqnarray*}

\begin{eqnarray*}
 & <\hat{P}_{A}(t),\hat{P}_{B}(t)>_{\rm v}=(\frac{\lambda_{Q}}{\Omega})^{2}\frac{1}{4\pi^{2}}
 \int d\omega \,\omega \, \frac{\sin\omega L}{\omega L}\cdot\\
 & |-i\omega(M_{1}-M_{2})e^{-i\omega t}+((i\Omega-\gamma_Q^{})M_{2}e^{i\Omega t}-(-i\Omega-\gamma_Q^{})M_{1}e^{-i\Omega t})e^{-\gamma_Q^{} t}|^{2}.\end{eqnarray*}

Physically, the v-part of the zeroth order correlators $<...>_{\rm v}$
effectively measure the response of the oscillator to vacuum fluctuations of the field.
The similarity between the integrands of $<\hat{Q}(t),\hat{Q}(t)>_{\rm v}$ in Eq.$(\ref{QQv})$ and
$<\hat{Q}_{A}(t),\hat{Q}_{B}(t)>_{\rm v}$ in Eq.$(\ref{QAQBv})$ is not surprising. 
In Appendix \ref{w2IHO} we show that, if we have
two inertial oscillators C and D in free space at a distance $L$ apart,  
with the same coupling constants but in opposite signs,
then $(\hat{Q}_{C}(t)+\hat{Q}_{D}(t))/2$ obeys the same equation of motion
as the one for $\hat{Q}(t)$ in our model. Thus we see that
the self correlator $<\hat{Q}(t),\hat{Q}(t)>_{\rm v}$ here
has the same value as $(<\hat{Q}_{C},\hat{Q}_{C}>+<\hat{Q}_{D},\hat{Q}_{D}>)/4
+<\hat{Q}_{C},\hat{Q}_{D}>/2$ at $t$ and
contains the part of correlations of vacuum fluctuation in free space
which is odd with respect to the $z_3 =0$ plane, whereas
$<\hat{Q}_{C}(t),\hat{Q}_{D}(t)>_{\rm v} \propto (-\lambda_Q)\lambda_Q$ here has exactly the same
value of $- <\hat{Q}_{A}(t),\hat{Q}_{B}(t)>_{\rm v} $ in \cite{LH_2IHO}
because the two oscillators in \cite{LH_2IHO} are identically coupled to the field
and so $<\hat{Q}_{A},\hat{Q}_{B}>_{\rm v} \propto \lambda_Q\lambda_Q$.

\subsection{Early-time Dynamics of Oscillator-field Entanglement}

\begin{figure}
\includegraphics[width=7cm]{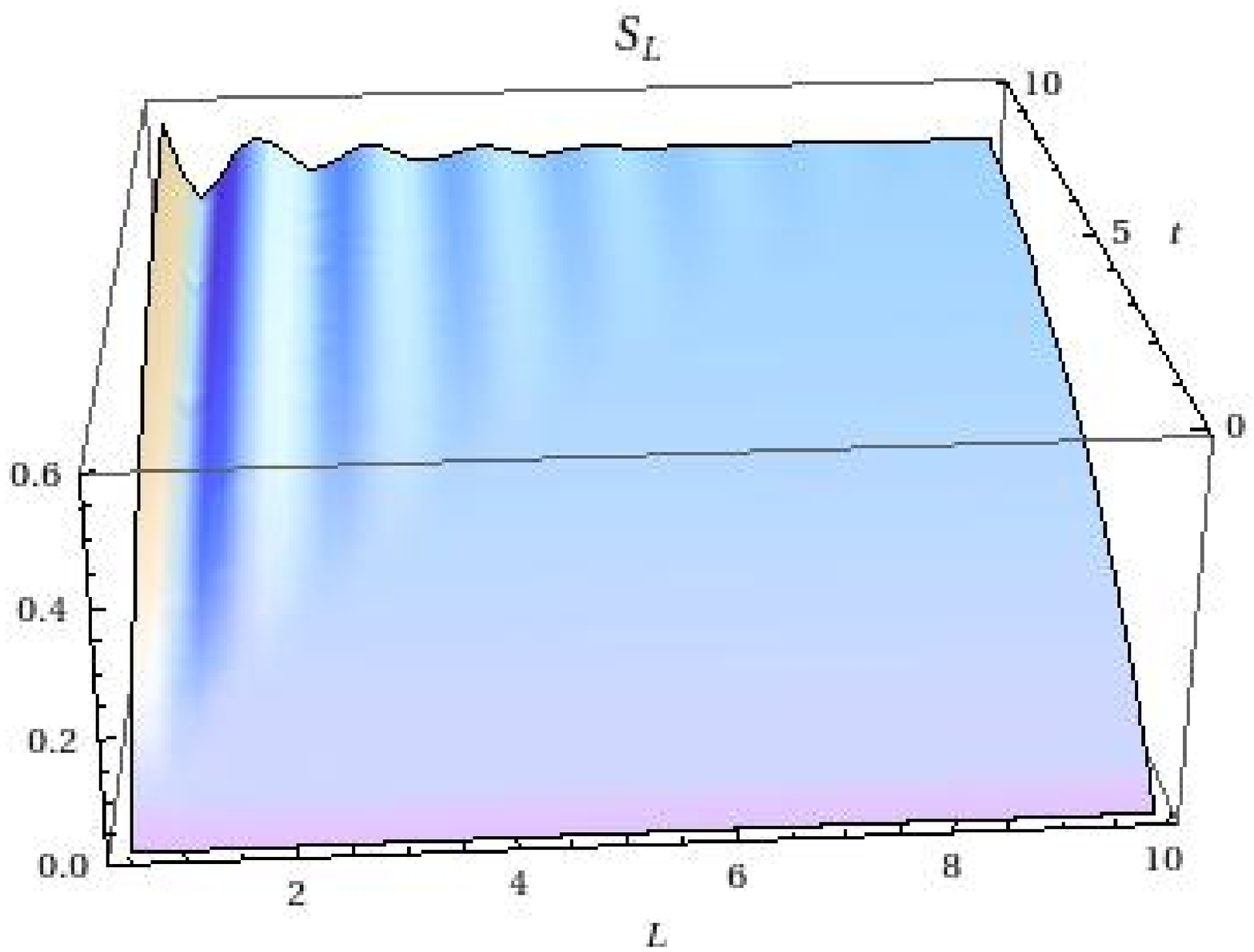}
\includegraphics[width=5cm]{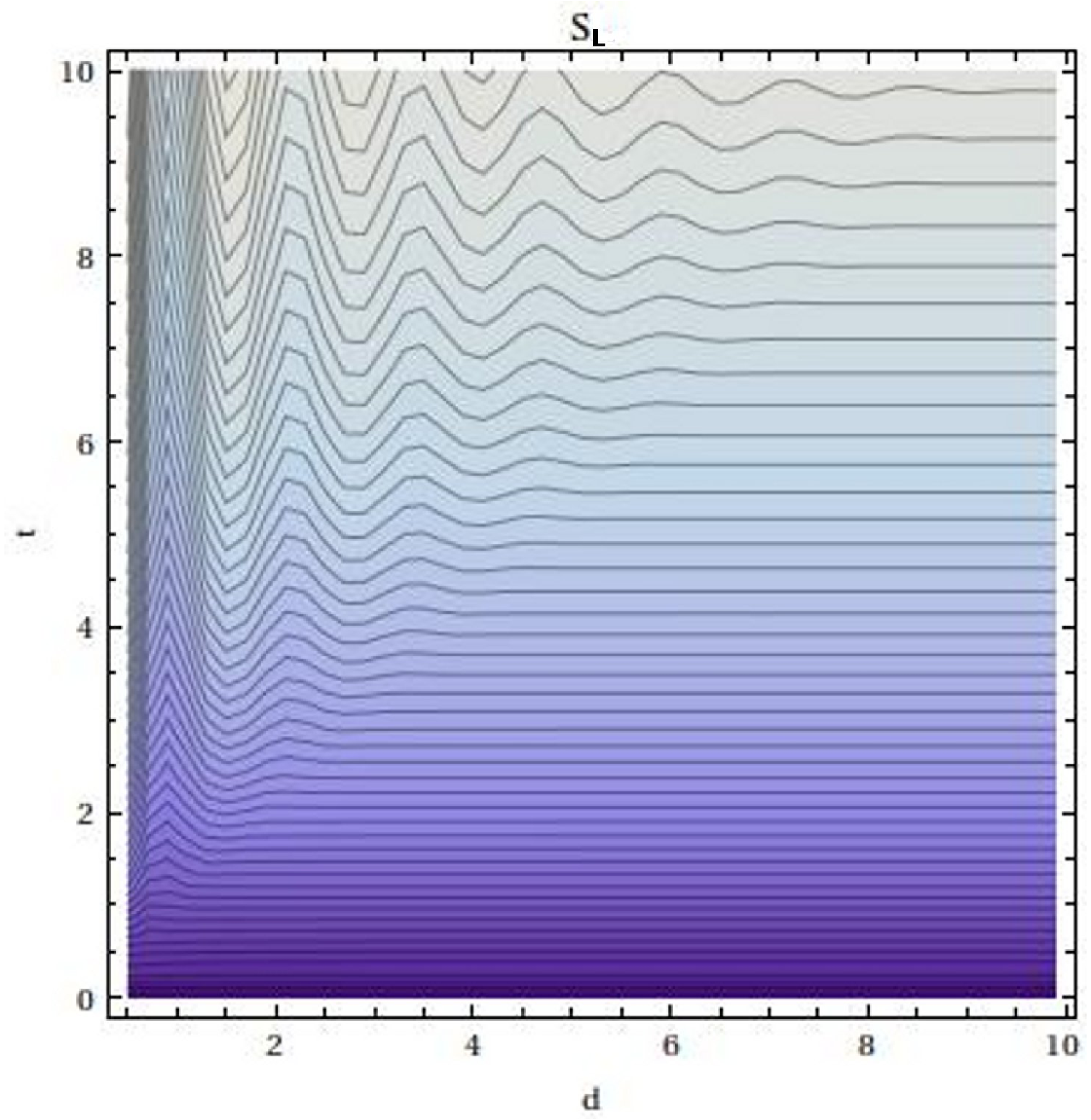}\\
\includegraphics[width=6cm]{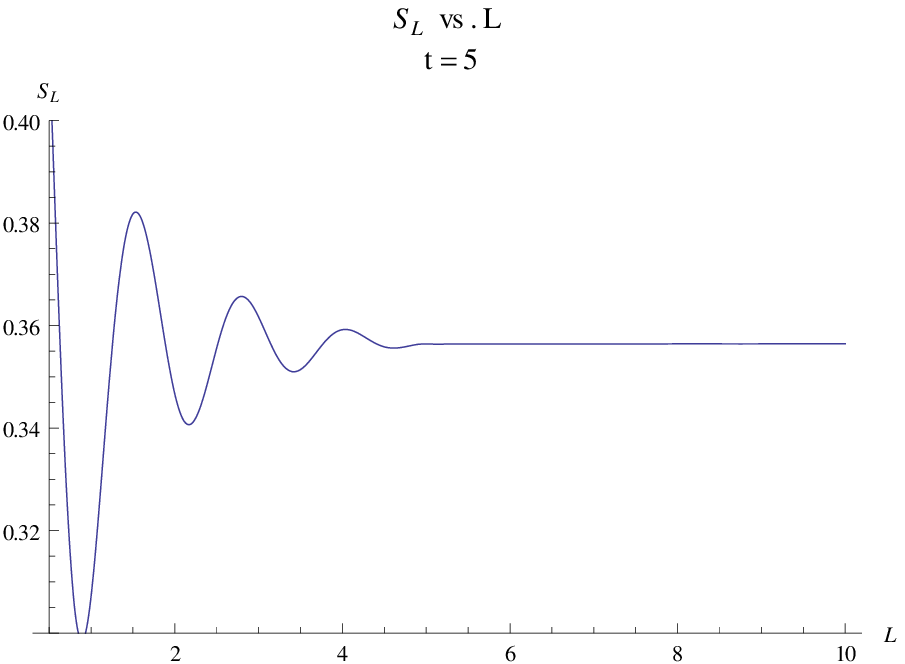}
\includegraphics[width=6cm]{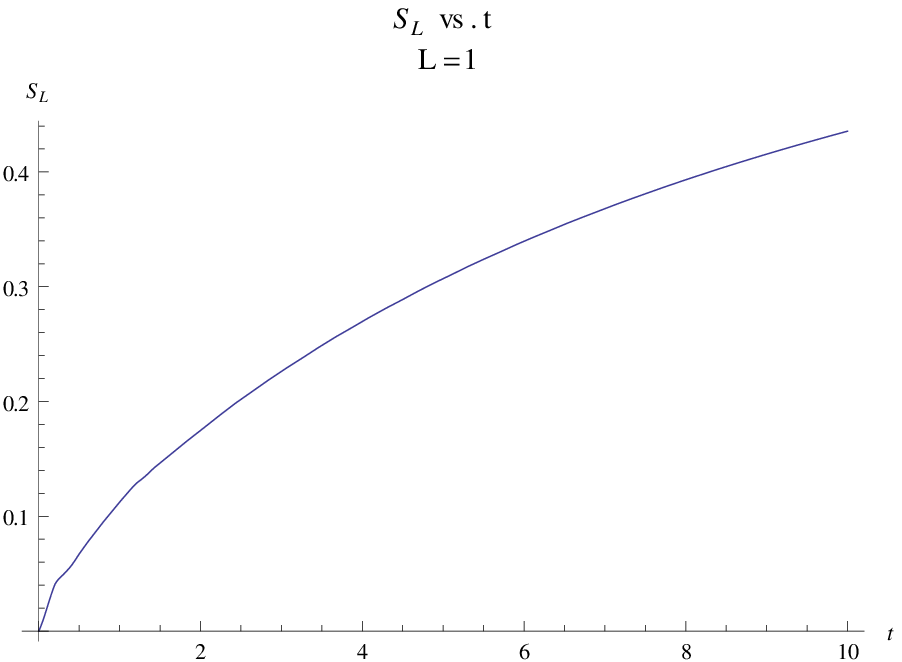}
\caption{Entanglement dynamics at early time where only zeroth order correlations
contribute.
Here $\gamma_Q^{}=0.02$ and $\Omega=5$. The upper left plot shows the linear entropy as a function of L and t, with
the upperright plot being its contour plot.
The lowerleft plot shows the dependence of linear entropy on L at a given instant of time whereas the lower right
plot exhibits how the linear entropy evolves with time for oscillator located at a certain distance.
}
\label{EntDynamics}
\end{figure}

With the previous results we can now investigate the evolution of oscillator-field
entanglement as the distance between the oscillator and the mirror changes.
Whereas it may be possible to obtain an approximate analytical expression
in the weak coupling limit, we can simply study the dependence of
the linear entropy on $L$ and $t$ numerically, as shown in Figure \ref{EntDynamics}.

Our numerical results reveal the following behaviors:

\begin{itemize}
\item For a given $L$, the entanglement between the oscillator and the field
increases monotonically at early times, 
showing that the oscillator is getting
more and more entangled with the field after the interaction is turned on.
(See Figure \ref{EntDynamics} (upper row) and (lower right).)
\item Within the light cone, 
at every given time in this stage, 
the oscillator-field entanglement
exhibits an oscillatory behavior with spatial frequency being $\Omega$.
This shows
that the growth rate of $S_L$
with which the oscillator gets entangled
with the field also oscillates as a function of $L$, as shown in
Figure \ref{EntDynamics} (upper row) and (lower left).
\end{itemize}
Such oscillatory behavior in $L$ is solely due to the oscillatory behavior
of the zeroth order correlators corresponding to $< \hat{R}_{C}, \hat{R}_{D}>_{\rm v}$,
$R=Q, P$, given in the last paragraph of the previous subsection.
After $\cos\theta$ in $k_3\equiv \omega\cos\theta$ has been integrated over the interval $[0, 1]$ in the integration
corresponding to $<\hat{Q}_{C},\hat{Q}_{D}>_{\rm v}$ in $(\ref{QQv})$ or $(\ref{QAQBv})$,
the field modes on resonance with the oscillator have constructive (destructive) interference
at the local maxima (minima) of $S_L$ against $L$ at fixed $t$ at early times.

In \cite{LH_2IHO} the early time dynamics of entanglement between the two oscillators exhibit
similar oscillatory dependence on the distance between them, which
is also due to distance-dependent correlations of vacuum fluctutaions
experienced by the oscillators.
More precisely, from $(\ref{QAQBv})$ one can see that the field modes with $\omega = n\pi/L$, $n=1,2,3,\cdots$,
has no contribution at all to the integration of $<\hat{Q}_{A},\hat{Q}_{B}>_{\rm v}$, and those
satisfying $\tan \omega L = \omega L$ give the maximum and minimum values of the factor
$\sin \omega L/ (\omega L)$ in the integrand. In the weak coupling limit the integration in
$(\ref{QAQBv})$ is mainly contributed by the poles at $\omega \approx \pm \Omega$, namely,
those mode on resonance with the oscillators. So $<\hat{Q}_{C},\hat{Q}_{D}>_{\rm v} \approx 0$
for $L \approx n \pi/\Omega$, while $<\hat{Q}_{C},\hat{Q}_{D}>_{\rm v}$ has local maximum or
minimum values when $L$ is about the solution of $\tan \Omega L = \Omega L$ in the weak coupling limit.

\section{Late-Time Stationary Limit of Oscillator-Field Entanglement}

\subsection{Quantum Langevin Equation and Covariance Matrix at Late-times}

One can compare the dynamics of the oscillator under the influence of the field
to the well-studied quantum Brownian motion (QBM) model, namely,
the retarded and Hadamard Green functions correspond to the dissipation
and noise kernels respectively and the function $I(\omega;{\bf x},{\bf y})$
represents the spectral density {[}see, e.g., \cite{HPZ,HM94}) and
the references therein.{]} In the same vein Eq. (\ref{FI-AtomEqn})
can equivalently be written as a quantum Langevin equation
\begin{equation}
M_{Q}\ddot{\hat{Q}}(t)+M_{Q}\Omega^{2}\hat{Q}(t)-\int_{t_{i}}^{t}d\tau\mu(t,\tau)\cdot\hat{Q}(\tau)=\hat{\xi}(t),\label{QBM-type-Eqn}\end{equation}
 in which
\begin{equation}
\mu(t,s)=\lambda_{Q}^{2}\ \theta(t-s)\int_{0}^{\Lambda}d\omega I(\omega;{\bf x}_Q^{},{\bf x}_Q^{})\sin\omega(t-s),\label{eq:dissipationkernel}\end{equation}
 and $\hat{\xi}(t)=\lambda_{Q}{\hat{\Phi}}_{0}({\bf x}_Q^{},t)$. In
the integrals over the frequency above and below we have assumed a
high frequency finite cutoff $\Lambda$ for the quantum field $\Lambda$
which regularizes the quantum field's retarded Green's function from
which we obtain the effective equations of motion of the oscillator \cite{LH_2IHO}.
In QBM language $\mu$ is the dissipation kernel, $\nu(t,s)$ is the
noise kernel  which quantifies the two-time correlation
of the Langevin forcing term $\hat{\xi}(t)$,
\begin{equation}
\nu(t,s)\equiv\langle\{\hat{\xi}(t),\hat{\xi}(s)\}\rangle=\lambda_{Q}^{2}\int_{0}^{\Lambda}d\omega I(\omega;{\bf x}_Q^{},{\bf x}_Q^{})\cos\omega(t-s).\label{eq:noisekernel}\end{equation}

 {[}From now on we simply denote $I(\omega)\triangleq I(\omega;{\bf x}_Q^{},{\bf x}_Q^{})$.{]}
Here the average is taken with respect to the initial state density matrix of the field.

By introducing the damping kernel $\gamma(t,s)$ defined by
\begin{equation}
\mu(\tau,s)=M_{Q}\frac{\partial}{\partial\tau}\gamma(\tau,s)=M_{Q}\frac{\partial}{\partial s}\gamma(\tau,s),\label{eq:dampingkernel}\end{equation}
 we can bring (\ref{FI-AtomEqn}) to the final form
\begin{equation}
M_{Q}\ddot{\hat{Q}}(t)+M_{Q}\Omega_{r}^{2}\hat{Q}(t)+M_{Q}\int_{t_{i}}^{t}d\tau\gamma(t,\tau)\cdot\dot{\hat{Q}}(\tau)+2M_{Q}\gamma(t,t_{i})\cdot\hat{Q}(t_{i})=\hat{\xi}(t),\label{eq:Langevin}\end{equation}
 where $\Omega_{r}^{2}=\Omega^{2}-2M_{Q}\gamma(t,t)$ is the renormalized
frequency.

The general solution to (\ref{QBM-type-Eqn}) is given by
\begin{equation}
\hat{Q}(t)=\hat{Q}_{0}(t)+\int_{t_{i}}^{t}ds\ \tilde{G}(t,s)\hat{\xi}(s),\label{Soln}\end{equation}
 where $\tilde{G}(t,s)$ is the retarded Green's function for (\ref{QBM-type-Eqn}) satisfying

\begin{equation}
M_{Q}\ddot{\tilde{G}}(t,s)+M_{Q}\Omega^{2}\tilde{G}(t,s)-\int_{t_{i}}^{t}d\tau\mu(t,\tau)\cdot\tilde{G}(\tau,s)=\delta(t-s).\label{GreenF}
\end{equation}

 For here and below all the Fourier components are defined for positive
frequency only, which correspond to Fourier transformation of function
defined on $t>0$. Accordingly, the Fourier space representation of
the above Green's function can be written as:
\begin{align}
\tilde{G}(\omega)\equiv & (M_{Q}(-\omega^{2}+\Omega^{2})-\mu(\omega))^{-1}\nonumber \\
\triangleq & (-\omega^{2}-i\omega\gamma(\omega)+\Omega_{r}^{2})^{-1}M_{Q}^{-1}.\end{align}

In the late-time stationary limit, because $\mu(t,s)$ leads to dissipation
of the oscillator's free motion, we see that
\begin{equation}
\tilde{\hat{Q}}(\omega)\to\tilde{G}(\omega)\tilde{\hat{\xi}}(\omega).\label{ }\end{equation}
 As has been explained before, if we assume the initial state of our
combined system to be Gaussian, then because the total Hamiltonian
is quadratic, the quantum state of the oscillator will always be Gaussian
and can be fully characterized by the covariance matrix
\begin{equation}
(\boldsymbol{V})_{ij}(t)=\frac{1}{2}\langle\hat{O}_{i}(t)\hat{O}(t)_{j}+\hat{O}_{j}(t)\hat{O}_{i}(t)\rangle,\end{equation}
 in which $\boldsymbol{\hat{O}}=(\hat{Q},\hat{P})$.

In the late-time stationary limit the elements $V_{QP}^{\infty}$
vanish, as can be inferred from $V_{QP}(t)=M\dot{V}_{QQ}(t)/2$ if
$V_{QQ}(t)$ approaches an asymptotic constant value.

The remaining non-zero elements of the covariance matrix at late times
are \cite{QOS}:

\begin{align}
V_{QQ}^{\infty}=\left. < \hat{Q}(t),\hat{Q}(t) > \right|_{t\to\infty} = & \int_{0}^{\Lambda}d\omega\ \tilde{G}^{*}(\omega)\cdot I(\omega)\cdot\tilde{G}(\omega),\\
V_{PP}^{\infty}=\left. < \hat{P}(t),\hat{P}(t) > \right|_{t\to\infty}= & \int d\omega\ \omega^{2}M_{Q}^{2}\tilde{G}^{*}(\omega)\cdot I(\omega)\cdot\tilde{G}(\omega).\end{align}

By applying the fluctuation-dissipation theorem \cite{FeynVern,HPZ}, one can eliminate
the noise kernel in the covariance matrix elements, giving

\begin{equation}
V_{QQ}^{\infty}=\frac{1}{\pi}\int_{0}^{\infty}d\omega{\rm Im}[\tilde{G}(\omega)],\end{equation}

\begin{equation}
V_{PP}^{\infty}=\frac{1}{\pi}\int_{0}^{\infty}d\omega\ \omega^{2}M_{Q}^{2}{\rm Im}[\tilde{G}(\omega)].\end{equation}

For detailed derivation leading to the above results, please refer
to Appendix \ref{LateCV}.

\subsection{Entanglement between Oscillator and Field in Free Space}

We will begin by calculating the late-time oscillator-field entanglement
for an oscillator in free space. We then compute the oscillator-field entanglement
with a mirror present and compare their differences. The first step
is to regularize the retarded Green's function of the field at the
trajectory of the detector. For an oscillator interacting with a massless
scalar field in free space, the entire system is governed by the action

\begin{equation}
S[Q,\dot{Q};\Phi,\partial_{\mu}\Phi]=\frac{1}{2}M_{Q}\int dt\ (\dot{Q}^{2}-\Omega^{2}Q^{2})+\frac{1}{2}\int dt\int d^{3}x\ \partial_{\mu}\Phi\partial^{\mu}\Phi+\lambda_{Q}\int dt\ Q(t)\Phi({\bf x}_Q^{},t).\end{equation}
 which is the same as Eq. (\ref{eq:Action_Half}) but without the
$x_{3}>0$ restriction.

Following \cite{LH_2IHO} we obtain

\begin{equation}
\left(\partial_{t}^{2}+2\gamma_{Q}^{}\partial_{t}+\Omega_{r}^{2}\right)\hat{Q}(t)=\lambda_{Q}\hat{\Phi}(t,{\bf x}_Q^{})/M_{Q},\end{equation}
 where 
$\Omega_{r}$ is the renormalized natural frequency.

For free space the retarded Green's function for (\ref{QBM-type-Eqn})
is given by
\begin{equation}
\tilde{G}(\omega)=\frac{1}{M_{Q}}[-(\omega^{2}+i\gamma_Q^{})^{2}+\tilde{\Omega}_{r}^{2}]^{-1},\end{equation}
 where $\tilde{\Omega}_{r}\equiv\Omega_{r}^{2}-\gamma^{2}$ from which
the late-time covariances can be computed:
\begin{eqnarray}
V_{QQ,{\rm free}}^{\infty} & = & \frac{1}{\pi}\int_{0}^{\Lambda}d\omega{\rm Im}\left[\frac{1}{M_{Q}}\frac{1}{-(\omega+i\gamma_Q^{})^{2}+\tilde{\Omega}_{r}^{2}}\right]\\
 & = & \frac{i}{2\pi M_{Q}\tilde{\Omega}_{r}}\ln\frac{\gamma_Q^{}-i\tilde{\Omega}_{r}}{\gamma_Q^{}+i\tilde{\Omega}_{r}},\nonumber \end{eqnarray}

\begin{eqnarray}
V_{PP,{\rm free}}^{\infty} & = & \frac{1}{\pi}\int_{0}^{\Lambda}d\omega{\rm Im}\left[M_{Q}\frac{\omega^{2}}{-(\omega+i\gamma_Q^{})^{2}+\tilde{\Omega}_{r}^{2}}\right]\\
 & = & M_{Q}\left\{ \frac{i}{2\pi\tilde{\Omega}_{r}}(\tilde{\Omega}_{r}^{2}-\gamma_Q^{2})\ln\frac{\gamma_Q^{}-i\tilde{\Omega}_{r}}{\gamma_Q^{}+i\tilde{\Omega}_{r}}+\frac{\gamma_Q^{}}{\pi}\left[2\Lambda-\ln\bigg(1+\frac{\gamma_Q^{2}}{\tilde{\Omega}_{r}^{2}}\bigg)\right]\right\} .\nonumber \end{eqnarray}
 Perturbatively in $\gamma_Q^{}$ one has

\begin{equation}
V_{QQ,{\rm free}}^{\infty}=\frac{1}{2M_{Q}\tilde{\Omega}_{r}}\left(1-\frac{2\gamma_Q^{}}{\pi\tilde{\Omega}_{r}}\right),\end{equation}

\begin{equation}
V_{PP,{\rm free}}^{\infty}=M_{Q}\left(\frac{\tilde{\Omega}_{r}}{2}+\frac{1}{\pi}\gamma_Q^{}\left[2(\ln\Lambda-\ln\tilde{\Omega}_{r})-\frac{\tilde{\Omega}_{r}^{2}}{\Lambda^{2}}-1\right]\right),\end{equation}
 which recovers the results in \cite{LH_2IHO}.

\subsection{Entanglement between Oscillator and Field under Mirror Reflection}

In the presence of a perfect mirror, the entire system is governed
by action (\ref{eq:Action_Half}). The Heisenberg equation of motion
for the oscillator after the same regularization as above is
\begin{equation}
\left(\partial_{t}^{2}+2\gamma_Q^{}\partial_{t}+\Omega_{r}^{2}\right)\hat{Q}(t)=-\frac{2\gamma_Q^{}}{4\pi L}\theta(t-L)\hat{Q}(t-L)+\lambda_{Q}\hat{\Phi}(t,{\bf x}_Q^{}),\label{RegEOM_Half}
\end{equation}
and correspondingly\begin{equation}
\tilde{G}(\omega)=\frac{1}{M_{Q}}[-(\omega+i\gamma_Q^{})^{2}+\tilde{\Omega}_{r}^{2}+(2\gamma_Q^{}e^{i\omega L}/L)]^{-1}.\end{equation}
 where the last term inside the square brackets shows the difference
from the free space results. It can be interpreted as the contribution
from the oscillator's mirror image located at a vertical distance $L/2$ behind
the mirror. 

For this case the exact late-time covariance matrix becomes
\begin{eqnarray}
V_{QQ,{\rm half-space}}^{\infty} & = & \frac{1}{\pi M_{Q}}\int_{0}^{\Lambda}d\omega{\rm Im}\left[%\frac{1}{M_{Q}}
\frac{1}{-(\omega+i\gamma_Q^{})^{2}+\tilde{\Omega}_{r}^{2}+(2\gamma_Q^{}e^{i\omega L}/L)}\right],\label{eq:Vxx} \\
V_{PP,{\rm half-space}}^{\infty} & = & \frac{M_{Q}}{\pi}\int_{0}^{\Lambda}d\omega{\rm Im}\left[%M_{Q}
\frac{\omega^{2}}{-(\omega+i\gamma_Q^{})^{2}+\tilde{\Omega}_{r}^{2}+(2\gamma_Q^{}e^{i\omega L}/L)}\right].\label{eq:Vpp}\end{eqnarray}

Assuming that the oscillator is only weakly coupled to the field, we can
perturbatively expand the above integrals and get

\begin{eqnarray}
V_{QQ,{\rm half-space}}^{\infty} & = & V_{QQ,{\rm free}}^{\infty} +\delta V_{QQ}^{\infty}+
O(\gamma_{Q}^{2}),\\
V_{PP,{\rm half-space}}^{\infty} & = & V_{PP,{\rm free}}^{\infty} + \delta V_{PP}^{\infty}+O(\gamma_{Q}^{2}),
\end{eqnarray}

 where the terms $\delta V_{QQ}^{\infty}$ and $\delta V_{PP}^{\infty}$
represent the first corrections to the covariance matrix elements
due to the presence of the mirror. Physically keeping only these terms
for a single reflection in this perturbative expansion is equivalent
to ignoring the multiple reflections between the oscillator and the mirror.
The exact form for the leading order correction is given below:
\begin{align}
\delta V_{QQ}^{\infty} & \triangleq\frac{1}{\pi M_{Q}}\int_{0}^{\Lambda}d\omega%\frac{1}{M_{Q}}
{\rm Im}\left[\frac{1}{-(\omega+i\gamma_Q^{})^{2}+\tilde{\Omega}_{r}^{2}}\left(\frac{-2\gamma_Q^{}e^{i\omega L}/L}{-(\omega+i\gamma_Q^{})^{2}+\tilde{\Omega}_{r}^{2}}\right)\right]\nonumber\\
 & =-\frac{1}{\pi}\frac{1}{M_{Q}\Omega_{r}}\frac{\gamma_Q^{}}{L}{\rm Re}\left[\left(i\frac{1}{\Omega_{r}^{2}}+\frac{L}{\Omega_{r}}\right)e^{i\Omega_{r}L}\Gamma[0,i\Omega_{r}L]\right],
 \label{dVxx}
\end{align}
\begin{eqnarray}
\delta V_{PP}^{\infty} & \triangleq & \frac{M_{Q}}{\pi}\int_{0}^{\Lambda}d\omega {\rm Im}\left[\frac{\omega^{2}}{-(\omega+i\gamma_Q^{})^{2}+\tilde{\Omega}_{r}^{2}}\left(\frac{-2\gamma_Q^{}e^{i\omega L}/L}{-(\omega+i\gamma_Q^{})^{2}+\tilde{\Omega}_{r}^{2}}\right)\right]\nonumber\\
 & = & -\frac{M_{Q}^{}\gamma_Q^{}}{\pi\Omega_{r}L}{\rm Re}\left[\left(-i+L\Omega_{r})\right)
 e^{i\Omega_{r}L}\Gamma[0,i\Omega_{r}L]\right],
\label{dVpp} %\nonumber
\end{eqnarray}
 in the limit of large cutoff $\Lambda$.

\begin{figure}
\includegraphics[width=8cm]{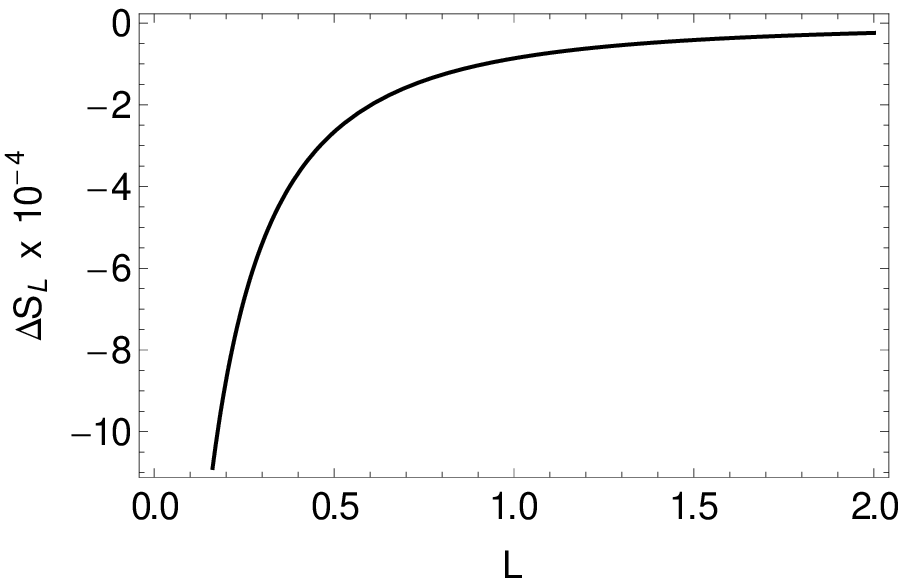}
\includegraphics[width=8cm]{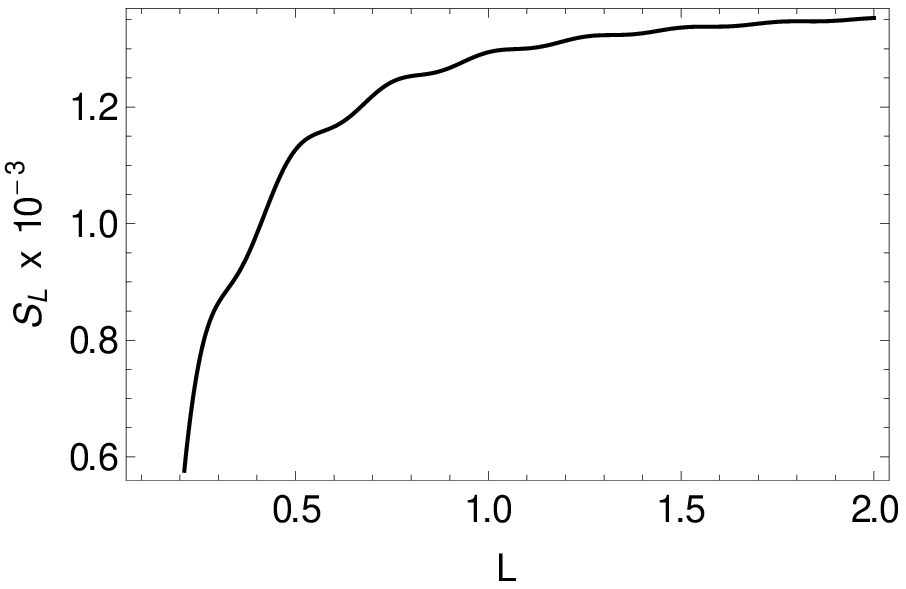} \\
\includegraphics[width=8cm]{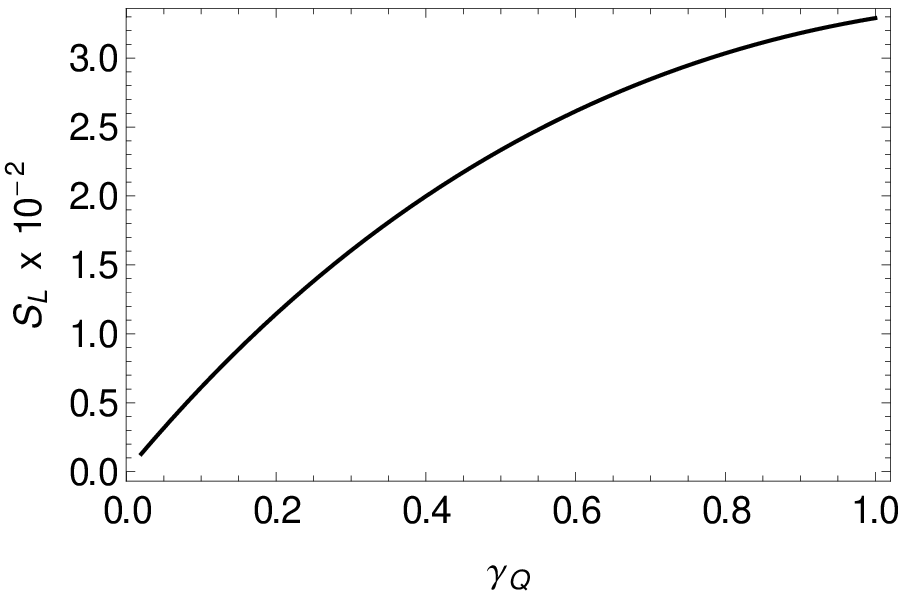}
\caption{(Upper-left) Leading-order corrections to linear entropy given in $(\ref{eq:LEntropy})$
as a function of distance between the
oscillator and its image due to the presence of the mirror.
(Upper-right) Linear entropy of the oscillator as a function of distance between the oscillator
and its image according to Eqs (\ref{eq:Vxx}) and (\ref{eq:Vpp}).
(Lower) Linear entropy as a function of $\gamma_Q$.
Here $M_{Q}=1$, $\gamma_Q^{}=0.02$, $\Omega_{r}=5$.}
\label{LinearEntropy}
\end{figure}

The change of linear entropy due to the presence of the mirror, compared
to the case of free space, is given as

\begin{equation}
\Delta S_{L}
\equiv S_{L,\,{\rm half-space}} - S_{L,\,{\rm free}}= 
=-\frac{2}{\pi}\frac{\gamma_Q^{}}{\Omega_{r}}{\rm Re}\left[e^{i\Omega_{r}L}\Gamma[0,i\Omega_{r}L]\right].\label{eq:LEntropy}
\end{equation}
In Figure \ref{LinearEntropy} (upper-left) we see that $\Delta S_L < 0$, thus
the presence of the mirror always acts to reduce the linear entropy between the oscillator and the field,
thereby causing them to be less entangled.

In the same plot one can also see that $\Delta S_{L}$ increases monotonically with $L$ and goes to $0$ as $L \to \infty$.

This fact can be intuitively understood by inspecting the Heisenberg
equation of motion of the oscillator's internal degree of freedom. As shown
in (\ref{RegEOM_Half}), due to mirror reflection, the oscillator's internal
degree of freedom will have an negative influence upon itself after
time L through field propagation. This will effectively reduce the
quadrature $V_{QQ}^{\infty}$ and $V_{PP}^{\infty}$ and hence cause
the oscillator to be less entangled with the field as it moves closer to
the mirror.

As one would naturally expect, as the oscillator becomes more and more strongly
coupled with the field, the oscillator-field entanglement increases monotonically,
as shown in the lower plot of Figure \ref{LinearEntropy}.

%%%%%%%%%%%%%%  new after 10pm Nov 25 %%%%%%%%%%%%%%%%%%%

\subsection{Effect of Mirror Image versus Effect of Real Object}

In our setup, the lowest order correction to self-correlators corresponds
to the physical process in which the oscillator emits a quanta and then
interact with it after it is reflected by the mirror. Ths process
contributes a correction of order $O(\gamma_Q^{}/L)$,
according to $(\ref{dVxx})$ and $(\ref{dVpp})$.
On the other hand, as stated in Appendix \ref{w2IHO},
in the case of two inertial oscillators with opposite coupling
constant, the lowest order correction to self-correlators corresponds
to the following physical process: oscillator A emits a quanta,
which interacts with oscillator B, then the back reaction
from oscillator B to the field echoes back and interacts with oscillator A.
The contribution of this process is of order $O(\gamma_Q^{2}/L^{2})$.
Therefore, the influence of the mirror and a real image oscillator correspond to
different physical processes and are of different orders in the coupling constant.

\section{Discussion}

We began with a rather naive question about 
whether quantum entanglement can exist between a physical object (the evil queen Q
modeled by a harmonic oscillator) with her mirror image. The immediate
answer from a formalist could be no, because an image is not a physical
object and only physical quantum objects can get entangled. What we
really wanted to find out in this inquiry is the effect of boundaries
on quantum entanglement, as that between an atom and its trap or cavity
surface. Our calculation leading to an answer to this query took on
three steps 1) Entanglement exists between an oscillator and a quantum field;
2) The presence of a mirror alters the field configuration, namely,
a perfectly reflecting mirror imposes a Dirichlet boundary conditions
on the field along the mirror surface; 3) The symmetry in the expression we derived for the entanglement between the oscillator and the field in half space with Dirichlet boundary conditions imposed on the mirror surface suggests that it is as if the oscillator was entangled with its image located at distance $L/2$ on the other side of the mirror. We expound the meaning of these statements with the following observations:
a) This entanglement is different both from that of an oscillator with the quantum field in free space, i.e.,  without a mirror, and that between two physical oscillators at distance $L$ apart. b) The statement of "entanglement between an oscillator with its image" should always be understood as "the entanglement of an oscillator with the quantum field in half space under Dirichlet boundary conditions". c) Note that this substitute description works only for Dirichlet conditions, it fails if a Neumann condition was imposed on the boundary instead.

Developing this theme further, one can see more clearly which parties are being entangled if one considers a microscopic model of a mirror such as the one considered by Galley et al \cite{OM-T1}, where the mirror's internal degrees of freedom is  modeled by an oscillator (called mirosc) with very light mass (whereby the quantum fluctuations of the mirrors internal degrees of freedom will be suppressed). With this setup we could then consider three physical degrees of freedom: the oscillator, the field and the mirosc (the light degrees of freedom making up the mirror). If one first considers the interaction between the mirosc and the field, this would yield in the zero mirosc mass  or infinite reflectivity limit the modified field configuration derived here. Then from the covariance matrix of the oscillator one can derive the entanglement between the oscillator and the modified field. With a microphysical model of the mirror one can calculate how the dynamics of the mirosc is altered while interacting with the field, how the field is modified, and how  the oscillator is entangled with the modified field. The advantage of this is that one can see clearly how this entanglement can be interpreted as the entanglement between the oscillator and the mirosc (for an example of the successive levels of coarse-graining , see, e.g., \cite{BehuninHu}). This reminds us of a similar procedure in electrostatics, namely, how the force between a charge and the induced surface charge density on a conducting plate can be calculated using the image charge method.
 
Another observation using elementary physics of wave reflection upon a mirror is  the following: Since the field configuration is at the base of inquires into boundary effects on the oscillator field entanglement, the behavior of reflected waves could provide some useful guide in building up our intuition on quantum entanglement in this setting.
Instead of a mirror with perfect reflectivity we can think of two
adjoining dielectric media 1, 2 with dielectric coefficients $\e_{1}<\e_{2}$
(the mirror situation considered above corresponds to the case where
$\e_{1}$ is the vacuum $\e_{0}$, much smaller than $\e_{2}$). For
waves propagating from a soft medium 1 to a hard medium 2, the reflected
wave is inverted. This show up in the reflected field configuration
carrying an opposite sign from the original field configuration, thus
partially canceling it. This cancelation effect is more severe near
the mirror surface and hence we see the decrease of entanglement as
the oscillator gets closer to the mirror. If this reasoning is correct then
in the reverse situation, if the queen were a water nymph living in
a lake (medium 2), looking up {at} the sky (the air is medium 1, where we have
assumed $\e_{2}>\e_{1}$). As waves from the heavy
medium entering a light medium will be reflected at the interface
with a positive amplitude, the entanglement would increase as the
nymph comes up close to the water-air surface. Thus we can add a
fourth factor, 4) that of parity in reflection. Fairy tales aside,
when the experimental techniques improve to the extent that one can
measure the quantum entanglement between an atom and the trap surface
these results could be of some practical value.

Our next paper \cite{ZBLH2} will treat the quantum entanglement between
an atom and a dielectric medium. We will adopt the influence functional
method recently used for the treatment of fluctuation forces between
an atom and a dielectric medium \cite{BehuninHu}. As a small corollary
we will be able to check on the correctness of the above qualitative
argument based on symmetry and parity considerations. Later papers
in this series will address entanglement domain and entanglement pattern,
and a parallel series on quantum entanglement in topologically non-trivial
spaces starting with $R^{1}\times S^{1}$ \cite{ZCLH1}, which can
be applied to atoms in a toroidal trap.\\

\noindent{\bf Acknowledgement} This work is supported in part by
the NSF Grant No. PHY-0801368 and the Nation Science Council of Taiwan under the Grant No. NSC 99-2112-M-018-001-MY3.
ROB is aided by an NSF-NSC U.S.-East Asia Ph.D student grant award to spend a summer in Taiwan in 2010. 

%%%%%%%%%%%%%%%%%%%%%%%%%%%%%%%%%%%%%%%%%%%%%%%%%%%%%%%%%%%%%%%%%%
\appendix

\section{Derivation of late-time covariance matrix}
\label{LateCV}
According to our definition, the full-time, exact expression for he
QQ-part of late time covariance matrix is

\begin{align}
V_{QQ}(t) & =\int_{0}^{\infty}d\omega\cdot I(\omega)\int_{0}^{t}d\tau_{1}\int_{0}^{t}d\tau_{2}\tilde{G}(\tau_{1})\cos[\omega(\tau_{1}-\tau_{2})]\tilde{G}(\tau_{2})\\
 & =\int_{0}^{\infty}d\omega\cdot I(\omega)\int_{0}^{t}d\tau_{2}\int_{\tau_{2}}^{\tau_{2}+t}d\bar{\tau}\tilde{G}(\bar{\tau}-\tau_{2})
\cos[\omega(\bar{\tau}-2\tau_{2})]\tilde{G}(\tau_{2}).
\end{align}

At late times the integral can be approximated by

\begin{align}
V_{QQ}(t) & \approx\int_{0}^{\infty}d\omega\cdot I(\omega)\int_{0}^{t}d\tau_{2}\int_{\tau_{2}}^{t}d\bar{\tau}G(\bar{\tau}-\tau_{2})\cos[\omega(\bar{\tau}-2\tau_{2})]G(\tau_{2})\\
 & =\int_{0}^{\infty}d\omega\cdot I(\omega)\int_{0}^{t}d\bar{\tau}\int_{0}^{\bar{\tau}}d\tau_{2}G(\bar{\tau}-\tau_{2})\cos[\omega(\bar{\tau}-2\tau_{2})]G(\tau_{2})\\
 & =\int_{0}^{\infty}d\omega\cdot I(\omega)\int_{0}^{t}d\tau Re\{[e^{-i\omega\tau}G(\tau)]*[e^{i\omega\tau}G(\tau)]\}.\end{align}

When $t\rightarrow+\infty$, the above approximated expressions become
exact, and we have

\begin{equation}
V_{QQ}^{\infty}=\int d\omega\ \tilde{G}^{*}(\omega)\cdot I(\omega)\cdot\tilde{G}(\omega),\end{equation}
 and similarly, the PP-part of the exact late-time covariance matrix
is

\begin{equation}
V_{PP}^{\infty}=\int d\omega\ \omega^{2}M_{Q}^{2}\tilde{G}^{*}(\omega)\cdot I(\omega)\cdot\tilde{G}(\omega).\end{equation}

According to previous definitions (\ref{eq:dissipationkernel}) (\ref{eq:dampingkernel})
we have

\begin{equation}
I(\omega)=\frac{2}{\pi}\omega M_{Q}Re[\gamma(\omega)],\end{equation}
 using the above formula we have \begin{align}
\tilde{G}^{*}(\omega)\cdot Re[\gamma(\omega)]\cdot\tilde{G}(\omega) & =\frac{1}{2}(\tilde{G}^{*}(\omega)\cdot\gamma^{*}(\omega)\cdot\tilde{G}^{T}(\omega)+\tilde{G}^{*}(\omega)\cdot\tilde{\gamma}(\omega)\cdot\tilde{G}^{T}(\omega))\\
 & =\frac{1}{2}(\frac{1}{2i\omega}(1-(-\omega^{2}+\Omega_{r}^{2})\tilde{G}^{*}(\omega)M_{Q})\cdot\tilde{G}^{T}(\omega)\nonumber \\
 & \;\; +\frac{1}{-2i\omega}\tilde{G}^{*}(\omega)\cdot(1-(-\omega^{2}+\Omega_{r}^{2})M_{Q}\tilde{G}^{T}(\omega))\nonumber \\
 & =\frac{1}{2\omega}Im[\tilde{G}(\omega)].\nonumber \end{align}

Therefore we find that, in the late-time limit, we can eliminate explicit
reference to the noise kernel and express the covariance matrix elements
as

\begin{equation}
V_{QQ}^{\infty}=\int d\omega\frac{1}{\pi}{\rm Im}[\tilde{G}(\omega)],\end{equation}

\begin{equation}
V_{PP}^{\infty}=\int d\omega\omega^{2}\frac{1}{\pi}M_{Q}^{2}{\rm Im}[\tilde{G}(\omega)].\end{equation}

\section{Comparison with the case of two inertial oscillators}
\label{w2IHO}

In Section VI of Ref. \cite{LH_2IHO}, we have obtained the late-time
correlators in the case with two identical Unruh-DeWitt detectors
at rest at $x_{3}=\pm L/2$. Let $\lambda_{0}\to-\lambda_{Q}$
for the left detector ($Q_{A}$) and $\lambda_{0}\to+\lambda_{Q}$
for the right detector ($Q_{B}$), one may wonder whether the detector
on the right ($Q_{B}$ at $x_{3}=+L/2$) in this two-detector case
would behave the same as the detector at the same position (say, $\tilde{Q}_{B}$
at $x_{3}=+L/2$) in the above single-detector case with its image
detector.

The answer is no. The presence of the other detector separated in
a distance $L$ from one detector introduces corrections to the late-time
correlators of a single detector, which are $O(\gamma_Q^{2}/L^{2})$
for the self correlators and $O(\gamma_Q^{}/L)$ for the cross correlators.
This is different from the above self correlators $V_{QQ}$ and $V_{PP}$,
which have $\delta V_{QQ}$ and $\delta V_{PP}$ in $O(\gamma_Q^{}/L)$.

This can be understood as follows. The late-time behavior of the correlators
in \cite{LH_2IHO} are determined by mode functions $q_{A}^{(+)}(t,{\bf k})$
and $q_{B}^{(+)}(t,{\bf k})$, whose equation of motion reads (Eq.(13)
in \cite{LH_2IHO} with $d$ and $\lambda_{0}$ modified)

\begin{eqnarray}
\left(\partial_{t}^{2}+2\gamma_Q^{}\partial_{t}+\Omega_{r}^{2}\right)q_{B}^{(+)}(t,{\bf k}) & = & -\frac{2\gamma_Q^{}}{ L}\theta(t-L)q_{A}^{(+)}(t-L,{\bf k})+\lambda_{Q}e^{-i\omega t+ik_{3}L/2},\label{eomqB}\\
\left(\partial_{t}^{2}+2\gamma_Q^{}\partial_{t}+\Omega_{r}^{2}\right)q_{A}^{(+)}(t,{\bf k}) & = & -\frac{2\gamma_Q^{}}{ L}\theta(t-L)q_{B}^{(+)}(t-L,{\bf k})-\lambda_{Q}e^{-i\omega t-ik_{3}L/2}.\label{eomqA}\end{eqnarray}

Similar to Eqs. $(A6)$-$(A11)$ in \cite{LH_2IHO}, these equations
give the same $c_{{\bf k}}^{0}$ for $q_{B}^{(+)}$ like $(A11)$,
while the counterpart for $Q_{A}$ is $-c_{{\bf k}}^{0}$. So the
late-time self correlators are still given by Eqs. $(48)$ and $(50)$
in \cite{LH_2IHO}, and the cross correlators are those in Eqs. $(49)$
and $(51)$ multiplied by $-1$. From Eqs. $(54)$ and $(55)$ in
\cite{LH_2IHO}, one can see that in weak coupling limit the late-time
cross correlators are $O(\gamma_Q^{}/L)$ and the correction to
the self correlators is $O(\gamma_Q^{2}/L^{2})$.

On the other hand, the detector in this paper
has
\begin{equation}
\left(\partial_{t}^{2}+2\gamma_Q^{}\partial_{t}+\Omega_{r}^{2}\right)q_{+}(t,{\bf k})=
-\frac{2\gamma_Q^{}}{ L}\theta(t-L)q_{+}(t-L,{\bf k})+{\lambda_{Q}\over M_Q}e^{-i\omega t}\sin\frac{k_{3}L}{2},\label{eomQ1B}
\end{equation}
where $q_{+}(t-L,{\bf k})$ on the right hand side can
be interpreted as the image of $q_{+}$. Indeed, $(\ref{eomQ1B})$ gives %($M_{Q}\equiv 1$)
\begin{equation}
\left. q_{+}(t,{\bf k})\right|_{t\gg1/\gamma_Q^{}}=\frac{-{\lambda_{Q}}e^{-i\omega t}\sin\frac{k_{3}L}{2}/M_Q}{\omega^{2}+2i\gamma_Q^{}\omega-\Omega_{r}^{2}-(2\gamma_Q^{}e^{i\omega L}/L)},
\end{equation}
so at late times,
\begin{eqnarray}
< \hat{Q}(t),\hat{Q}(t)>_{\rm v} 
&= & \hbar\int\frac{d^{3}k}{(2\pi)^{3}2\omega}\left| q_{+}(t,{\bf k})\right|^{2}\nonumber \\
 && \stackrel{t\to\infty}{\longrightarrow} 
 \frac{\hbar}{\pi M_Q}\int_{0}^{\infty}d\omega\,{\rm Im}\left[\frac{1}{\omega^{2}-2i\gamma_Q^{}\omega-\Omega_{r}^{2}-
 ( 2\gamma_Q^{} e^{-i\omega L}/L)}\right],
\end{eqnarray}
which is exactly the $V_{QQ}^{\infty}$ in $(\ref{eq:Vxx})$ after
letting $\hbar=1$. %M_{Q}=1$.

Note that the first order correction to $V_{QQ}^{\infty}$ in free
space has exactly the same value as the late-time cross correlator
$\left<\right.\{\hat{Q}_{A},\hat{Q}_{B}\}\left.\right>$ in the case
with two inertial detectors considered above. However, the latter
will not enter the reduced density matrix of $\hat{Q}_{B}$. Compare
$(\ref{eomQ1B})$ with $(\ref{eomqB})$ and $(\ref{eomqA})$, one
can see that it is $(q_{B}^{(+)}+q_{A}^{(+)})/(2i)$ rather than $q_{B}^{(+)}$
has the same late-time behavior as $q_{+}$ 
for $M_Q=1$.

%%%%%%%%%%%%%%%%%%%%%%%%%%%%%%%%%%%%%%%%%%%%%%%%%%%%%%%%%%%%%%%%%%%%


\begin{thebibliography}{26}

\bibitem{EntQFT} P. Calabrese and J. Cardy, ``Entanglement Entropy and
Quantum Field Theory'', J. Stat. Mech. (2004) P06002; H.Casini and M.Huerta, ``Entanglement entropy
in free quantum field theory'', J. Phys. A42:504007 (2009).

\bibitem{SCH1} K. Sinha, N. Cummings and B. L. Hu, {}``Protecting
and Dynamically Generating Entanglement in a Two-Atom Two-Field-Mode
Model\textquotedbl{} arXiv:1004.1834 (2010)


\bibitem{Sabrina} M. Scala, B. Militello, A. Messina, S. Maniscalco,
J. Piilo, and K. Suominen, J. of Phys. A 40, 14527 (2007).

\bibitem{Miao} H. Miao, S. Danilishin, and Y. Chen, Universal Quantum
Entanglement between an Oscillator and Continuous Fields, Phys. Rev.
A (2009).

\bibitem{Marshall} W. Marshall, C. Simon, R. Penrose, and D. Bouwmeester,
Towards Quantum Superpositions of a Mirror, Phys. Rev. Lett. 91, 130401
(2003).

\bibitem{ASH} C. Anastopoulos, S. Shresta and B. L. Hu, Quantum Information
Processing 8, 594 (2009), a summary of earlier unpublished work in
{[}arXiv:quant-ph/0610007{]}.

\bibitem{FCAH_2atom} Chris Fleming, Nicholas Cummings, C. Anastopoulos
and B. L. Hu, {}``Non-Markovian Dynamics and Entanglement of Two-level
Atoms in a Common Field\textquotedbl{} J. Phys. A (2012) {[}arXiv:1101.2668{]}

\bibitem{SCH2} K. Sinha, N. Cummings and B. L. Hu,{}``Effect of Interatomic
Separation on Entanglement Dynamics in a Two-Atom Two-Mode Model\textquotedbl{}
J. Phys. A (2012) arXiv:1108.2681

\bibitem{LH_2IHO} S. Y. Lin and B. L. Hu, {}``Temporal and Spatial
Dependence of Quantum Entanglement from Field Theory Perspective\textquotedbl{},
Phys. Rev. D 79, 085020 (2009) {[}arXiv:0812.4391{]}

\bibitem{NQBM} C. H. Fleming, B. L. Hu and A. Roura,{}``Quantum
Brownian motion of multipartite systems with entanglement dynamics\textquotedbl{}
arXiv:1106.5752

\bibitem{QOS} C. H. Fleming, B. L. Hu and A. Roura,{}``Exact analytical
solutions to the master equation of quantum Brownian motion for a
general environment\textquotedbl{}, Annals of Physics 326,1207 (2011)

\bibitem{ZFHed} Rong Zhou, Chris Fleming and B. L. Hu, {}``Entanglement
Domain\textquotedbl{} (in preparation)


\bibitem{Edomain} M.M. Wolf, F. Verstraete, M.B. Hastings and J.I. Cirac,
Area laws in quantum systems: mutual information
and correlations, Phys. Rev. Lett. 100, 070502 (2008);
J. Eisert, M. Cramer and M.B. Plenio, Area laws for
the entanglement entropy, Rev. Mod. Phys.
82, 277 (2010)


\bibitem{ZFHep} Rong Zhou, Chris Fleming and B. L. Hu, {}``Entanglement
Pattern\textquotedbl{} (in preparation)


\bibitem{Epattern} C. Holzhey, F. Larsen and F. Wilczek, ''Geometric
and Renormalized Entropy in Conformal Field Theory'',
Nucl. Phys. B424, 443 (1994). A. Kitaev and J. Preskill, ''Topological entangle-
ment entropy'', Phys. Rev. Lett. 96, 110404 (2006);
E. Fradkin and J. E. Moore, ``Entanglement entropy of
2D conformal quantum critical points: hearing the shape
of a quantum drum?'', Phys.Rev.Lett.97, 050404 (2006)


\bibitem{BehuninHu} R. Behunin and B. L. Hu, {}``Nonequilibrium
Atom-Dielectric Forces Mediated by a Quantum Field\textquotedbl{}
Phys. Rev. A 84, 012902 (2011) arxiv:quant-ph/11021765

\bibitem{NonEq QFT} E. Calzetta and B. L. Hu,\textsl{ Nonequilibrium
Quantum Field Theory}, Cambridge (2008)

\bibitem{FeynVern} R. P. Feynman and F. L. Vernon, Ann. Phys. (N.Y.)
\textbf{24}, 118 (1963)

\bibitem{CRV} E. Calzetta, A. Roura, and E. Verdaguer, Physica A
\textbf{319}, 188 (2003), arXiv:quant-ph/0011097.

\bibitem{RHA} A. Raval, B. L. Hu and J. Anglin, Phys. Rev. D \textbf{5}3,
7003 (1996)

\bibitem{RHK} A. Raval, B. L. Hu and D. Koks, Phys.Rev.D 55, 4795 (1997)

\bibitem{HPZ}B. L. Hu, Juan Pablo Paz, and Yuhong Zhang, Phys. Rev.
D 45, 2843 (1992)

\bibitem{HM94}B. L. Hu, A. Matacz, Phys. Rev.D 49, 6612 (1994)

\bibitem{OneModePurity}M. G. A. Paris, F. Illuminati, A. Serafini and
S.D e Siena, Phys. Rev.A \textbf{68}, 012314 (2003)

\bibitem{OneModeVEnt} Alessio Serafini, Fabrizio Illuminati and Silvio
De Siena, J. Phys. B: At. Mol. Opt. Phys. \textbf{37} (2004) L21-28

\bibitem{LogaNega} M. B. Plenio, Phys. Rev. Lett. \textbf{95}, 090503
(2005)

\bibitem{InhomoBook}W. C. Chew, {\sl Waves and Fields in Inhomogeneous Media},
IEEE Press (1995)

\bibitem{OM-T1} Chad Galley, Ryan Behunin and B. L. Hu, ``Theory of Optomechanics: Oscillator-Field Model of Moving Mirrors" (in preparation)

\bibitem{ZBLH2} Rong Zhou, R. Behunin, S. Y. Lin and B. L. Hu, {}``Atom-Dielectric
Entanglement\textquotedbl{} (in preparation)

\bibitem{ZCLH1} Rong Zhou, C. H. Chou, S. Y. Lin and B. L. Hu, in
preparation


\end{thebibliography}
\end{document}